\documentclass[12pt]{article}
\usepackage{amsmath,amssymb,multicol}
\usepackage{graphicx,psfrag,epsf}
\usepackage{enumerate}
\usepackage{natbib}
\usepackage{url} 
\usepackage{subfigure}
\newcommand{\blind}{0}

\def\xo#1{x_{({#1})}}

\def\tx{\tilde x}
\def\R{\mathbb{R}}

\def\hamise{h_{\mathrm{AMISE}}}

\usepackage[usenames,dvipsnames]{color}
\usepackage{xcolor}
\definecolor{col1}{HTML}{0072B2}
\definecolor{col2}{HTML}{000000}
\definecolor{col3}{HTML}{808080}
\definecolor{col4}{HTML}{e79F00}
\definecolor{col5}{HTML}{009E73}
\definecolor{col6}{HTML}{F0E442}
\definecolor{col7}{HTML}{660066}
\definecolor{col8}{HTML}{9ad0F3}
\definecolor{col9}{HTML}{D55E00}
\definecolor{col10}{HTML}{CC79A7}

\newtheorem{theorem}{Theorem}
\newtheorem{remark}[theorem]{Remark}

\begin{document}

\def\spacingset#1{\renewcommand{\baselinestretch}%
{#1}\small\normalsize} \spacingset{1}
\def\abs#1{\vert {#1} \vert}


\if0\blind
{
  \title{\bf Fast Exact Univariate Kernel Density Estimation}
  \author{David P. Hofmeyr\hspace{.2cm}\\
    Department of Statistics and Actuarial Science, Stellenbosch University
    }
  \maketitle
} \fi

\if1\blind
{
  \bigskip
  \bigskip
  \bigskip
  \begin{center}
    {\LARGE\bf Title}
\end{center}
  \medskip
} \fi

\bigskip
\begin{abstract}
This paper presents new methodology for computationally efficient kernel density estimation. It is shown that a large class of kernels allows for exact evaluation of the density estimates using simple recursions. The same methodology can be used to compute density derivative estimates exactly. Given an ordered sample the computational complexity is linear in the sample size.
Combining the proposed methodology with existing approximation methods results in extremely fast density estimation.
Extensive experimentation documents the effectiveness and efficiency of this approach compared with the existing state-of-the-art. 
\end{abstract}

\noindent%
\vfill

\newpage
\spacingset{1.45} 
\section{Introduction}
\label{sec:intro}

Estimation of density functions is a crucial task in exploratory data analysis, with broad application in the fields of statistics, machine learning and data science. Here a sample of observations, $x_1, ..., x_n$, is assumed to represent a collection of realisations of a random variable, $X$, with unknown density function, $f$. The task is to obtain an estimate of $f$ based on the sample values. Kernel based density estimation is arguably the most popular non-parametric approach. In kernel density estimation, the density estimator, $\hat f$, is given by a mixture model comprising a large number of (usually $n$) components. In the canonical form, one has
\begin{align}\label{eq:kde}
\hat f(x) = \frac{1}{nh}\sum_{i=1}^n K\left(\frac{x - x_i}{h}\right),
\end{align}
where $K(\cdot)$ is called the kernel, and is a density function in its own right, satisfying $K\geq 0$, $\int K = 1$.
The parameter $h > 0$ is called the bandwidth, and controls the smoothness of $\hat f$, with larger values resulting in a smoother estimator. A direct evalution of~(\ref{eq:kde}) at a collection of $m$ evaluation points, $\{\tx_1, ..., \tx_m\}$, has computational complexity $\mathcal{O}(nm)$, which quickly becomes prohibitive as the sample size becomes large, especially if the function estimate is required at a large number of evaluation points. Furthermore many popular methods for bandwidth selection necessitate evaluating the density estimate (or its derivatives) at the sample points themselves~\citep{scott1987biased,pw1976choice,sheather1991reliable}, making the procedure for choosing $h$ quadratic in computational complexity. Existing methods which overcome this quadratic complexity barrier are limited to kernels with bounded support and the Laplace kernel~\citep{fan1994fast}, or they rely on approximations. Popular approximations including binning~\citep{ScottS1985, HallW1994} and the fast Gauss~\cite[FGT]{YangDGD2003} and Fourier~\cite[FFT]{Silverman1982} tranforms, as well as combinations of these.  A more recent approach~\citep{RaykarDZ2010} relies on truncations of the Taylor series expansion of the kernel function. Generally speaking these methods reduce the complexity to $\mathcal{O}(n+m)$, with the constant term depending on the desired accuracy level.

In this paper the class of kernels of the form $K(x) = \mbox{poly}(|x|)\exp(-|x|)$, where $\mbox{poly}(\cdot)$ denotes a polynomial function of finite degree, 
is considered. It is shown that these kernels allow for extremely fast and exact evaluation of the corresponding density estimates. This is achieved by defining a collection of $\mathcal{O}((\alpha + 1) n)$ terms, where $\alpha$ is the degree of the polynomial, of which the values $\{\hat f(x_1), ..., \hat f(x_n)\}$ are linear combinations. These terms arise from exploiting the binomial expansion of polynomial terms and the trivial factorisation of the exponential function.  Furthermore these terms can be computed recursively from the order statistics of the sample. Given an ordered sample, the exact computation of the collection of values $\{\hat f(x_1), ..., \hat f(x_n)\}$ therefore has complexity $\mathcal{O}((\alpha + 1) n)$. Henceforth we will use $\mbox{poly}_\alpha(\cdot)$ to denote a polynomial function of degree $\alpha$.
 An important benefit of the proposed kernels over those used in the fast sum updating approach~\citep{fan1994fast}, is that bounded kernels cannot be reliably used in cross validation pseudo-likelihood computations. This is because the likelihood for points which do not lie within the support of the density estimate based on the remaining points is zero. Numerous popular bandwidth selection techniques can therefore not be applied.
%
\begin{remark}
The derivative of a $\mbox{poly}_{\alpha}(|x|)\exp(-|x|)$ function is equal to $x$ multiplied by a $\mbox{poly}_{\alpha-1}(|x|)\exp(-|x|)$ function, provided this derivative exists. The proposed methodology can therefore be used to exactly and efficiently evaluate $\{\hat f^{(k)}(x_1), ..., \hat f^{(k)}(x_n)\}$, where $\hat f^{(k)}$ denotes the $k$-th derivative of $\hat f$. Although a given $\mbox{poly}(|x|)\exp(-|x|)$ function is not infinitely differentiable at $0$, for a given value of $k$ it is straightforward to construct a $\mbox{poly}(|x|)\exp(-|x|)$ function with at least $k$ continuous derivatives. An alternative is to utilise leave-one-out estimates of the derivative, which can be computed for any $\mbox{poly}(|x|)\exp(-|x|)$ function provided no repeated values in the sample.
\end{remark}

\begin{remark}
The proposed class of kernels is extremely rich. The popular Gaussian kernel is a limit case, which can be seen by considering that the density of an arbitrary sum of Laplace random variables lies in this class.
\end{remark}


The remainder of the paper is organised as follows. In Section~\ref{sec:kernels} the kernels used in the proposed method are introduced, and relevant properties for kernel density estimation are discussed. It is shown that density estimation using this class of kernels can be performed in linear time from an ordered sample using the recursive formulae mentioned above. An extensive simulation study is documented in Section~\ref{sec:sim}, which shows the efficiency and effectiveness of the proposed approach. A final discussion is given in Section~\ref{sec:conclusions}.

\section{Computing Kernel Density Estimates Exactly}\label{sec:kernels}

This section is concerned with efficient evaluation of univariate kernel density estimates. A general approach for evaluating the estimated density based on kernels which are of the type $K(x) = \mbox{poly}(|x|)\exp(-|x|)$ is provided. These kernels admit a convenient algebraic expansion of their sums, which allows for the evaluation of the density estimates using a few simple recursions. The resulting computational complexity is $\mathcal{O}((\alpha + 1) n)$ for an ordered sample of size $n$, where $\alpha$ is the degree of the polynomial.

To illustrate the proposed approach we need only consider the evaluation of a function of the type
\begin{align}\label{eq:proxy.f}
\sum_{i=1}^n |x - x_i|^\alpha\exp\left(-\frac{|x - x_i|}{h}\right),
\end{align}
for an arbitrary $\alpha \in \{0, 1, 2, ...\}$. The extension to a linear combination of finitely many such functions, of which $\hat f$ is an example, is trivial. To that end let $\xo1 \leq \xo2 \leq ... \leq \xo n$ be the order statistics from the sample. Then define for each $k = 0, 1, 2, ..., \alpha$ and each $j = 0, 1, 2, ..., n$ the terms
\begin{align}
\ell(k, j) &= \sum_{i=1}^j(-\xo{i})^k\exp\left(\frac{\xo{i}-\xo{j}}{h}\right),\\
r(k, j) & = \sum_{i=j+1}^n (\xo{i})^k\exp\left(\frac{\xo{j}-\xo{i}}{h}\right),
\end{align}
where for convenience $\ell(k, 0)$ and $r(k, n)$ are set to zero for all $k$. Next, for a given $x \in \R$ define $n(x)$ to be the number of sample points less than or equal to $x$, i.e., $n(x) = \sum_{i=1}^n \delta_{x_i} ((-\infty, x])$, where $\delta_{x_i}(\cdot)$ is the Dirac measure for $x_i$. Then,
\begin{align*}
\sum_{i=1}^n |x -& x_i|^\alpha\exp\left(-\frac{|x - x_i|}{h}\right) = \sum_{i=1}^n |x - \xo i|^\alpha\exp\left(-\frac{|x - \xo i|}{h}\right)\\
& = \sum_{i=1}^{n(x)} (x - \xo i)^\alpha\exp\left(\frac{\xo i - x}{h}\right) + \sum_{i=n(x)+1}^n (\xo i - x)^\alpha\exp\left(\frac{x - \xo i}{h}\right)\\
&= \exp\left(\frac{\xo {n(x)}-x}{h}\right)\sum_{i=1}^{n(x)}\sum_{k=0}^\alpha {\alpha \choose k} x^{\alpha-k} (- \xo i)^k\exp\left(\frac{\xo i - \xo {n(x)}}{h}\right)\\
& \hspace{20pt} + \exp\left(\frac{x-\xo {n(x)}}{h}\right)\sum_{i=n(x)+1}^n\sum_{k=0}^\alpha {\alpha \choose k} \xo i^k (- x)^{\alpha-k}\exp\left(\frac{\xo {n(x)} - \xo i}{h}\right)\\
&= \sum_{k=0}^\alpha {\alpha \choose k}\left( \exp\left(\frac{\xo {n(x)}-x}{h}\right)x^{\alpha-k}\ell(k, n(x)) + \exp\left(\frac{x-\xo {n(x)}}{h}\right)(- x)^{\alpha-k}r(k, n(x))\right).
\end{align*}
Now, if $x$ is itself an element of the sample, say $x = \xo j$, then we have
\begin{align*}
\sum_{i=1}^n |\xo j - x_i|^\alpha \exp\left(-\frac{|\xo j - x_i|}{h}\right)  = \sum_{k=0}^\alpha {\alpha \choose k}\left((\xo j)^{\alpha-k} \ell(k,j) + (-\xo j)^{\alpha-k} r(k,j)\right).
\end{align*}
The values $\hat f(\xo 1), ..., \hat f(\xo n)$ can therefore be expressed as linear combinations of terms in $\bigcup_{k, j}\{\ell(k, j), r(k, j)\}$. Next it is shown that for each $k = 0, ..., \alpha$, the terms $\ell(k, j), r(k, j)$ can be obtained recursively. Consider,
\begin{align*}
\ell(k, j+1) &= \sum_{i=1}^{j+1}(-\xo{i})^k\exp\left(\frac{\xo{i}-\xo{j+1}}{h}\right)\\
&= \sum_{i=1}^{j}(-\xo{i})^k\exp\left(\frac{\xo{i}-\xo j + \xo j -\xo{j+1}}{h}\right) + (-\xo{j+1})^k\\
&= \exp\left(\frac{\xo{j}-\xo{j+1}}{h}\right)\sum_{i=1}^{j}(-\xo{i})^k\exp\left(\frac{\xo{i}-\xo{j}}{h}\right) + (-\xo{j+1})^k\\
&= \exp\left(\frac{\xo{j}-\xo{j+1}}{h}\right)\ell(k, j) + (-\xo{j+1})^k.
\end{align*}
And similarly,
\begin{align*}
r(k, j-1) &= \sum_{i=j}^n (\xo{i})^k\exp\left(\frac{\xo{j-1}-\xo{i}}{h}\right)\\
&= \exp\left(\frac{\xo{j-1}-\xo{j}}{h}\right)\left(\sum_{i=j+1}^n (\xo{i})^k\exp\left(\frac{\xo{j}-\xo{i}}{h}\right) + (\xo{j})^k\right)\\
&= \exp\left(\frac{\xo{j-1}-\xo{j}}{h}\right)\left(r(k,j) + (\xo{j})^k\right).
\end{align*}
The complete set of values $\hat f(\xo 1), ..., \hat f(\xo n)$ can thus be computed with a single forward and a single backward pass over the order statistics, requiring $\mathcal{O}((\alpha + 1) n)$ operations in total. On the other hand evaluation at an arbitrary collection of $m$ evaluation points requires $\mathcal{O}((\alpha + 1)(n+m))$ operations.

Relevant properties of the chosen class of kernels can be simply derived. Consider the kernel given by
\begin{align*}
K(x) = c\exp(-|x|)\sum_{k=0}^\alpha \beta_k|x|^k,
\end{align*}
where $c$ is the normalising constant. Of course $c$ can be incorporated directly into the coefficients $\beta_0, ..., \beta_\alpha$, but for completeness the un-normalised kernel formulation is also considered. It might be convenient to a practitioner to only be concerned with the shape of a kernel, which is defined by the relative values of the coefficients $\beta_0, ..., \beta_\alpha$, without necessarily concerning themselves initially with normalisation. Many important properties in relation to the field of kernel density estimation can be simply derived using the fact that
\begin{align*}
\int_{-\infty}^\infty |x|^k\exp(-|x|)dx = 2\int_0^\infty x^k\exp(-x)dx = 2k!
\end{align*}
Specifically, one has
\begin{align*}
c^{-1} &= \sum_{k=0}^\alpha \beta_k\int_{-\infty}^\infty |x|^k \exp(-|x|)dx = 2\sum_{k=0}^\alpha \beta_k k!\\
\sigma_K^2 &:= \int_{-\infty}^\infty x^2 K(x) dx = c\sum_{k=0}^\alpha\int_{-\infty}^\infty |x|^{k+2}\exp(-|x|)dx = 2c\sum_{k=0}^\alpha \beta_k (k+2)!\\
R(K) & := \int_{-\infty}^\infty K(x)^2 = c^2\sum_{k=0}^\alpha\sum_{j=0}^\alpha \beta_k\beta_j \int_{-\infty}^\infty |x|^{k+j}\exp(-2|x|)dx\\
&= c^2\sum_{k=0}^\alpha\sum_{j=0}^\alpha \beta_k\beta_j \frac{1}{2}\int_{-\infty}^\infty \frac{|x|^{k+j}}{2^{k+j}}\exp(-|x|)dx = c^2\sum_{k=0}^\alpha\sum_{j=0}^\alpha \frac{\beta_k\beta_j}{2^{k+j}}(k+j)!
\end{align*}
Furthermore it can be shown that for $K(x)$ to have at least $k$ continuous derivatives it is sufficient that
\begin{align*}
\beta_i \propto \frac{1}{i!}, \mbox{for all } i = 0, 1, ..., k.
\end{align*}
Choosing the simplest kernel from the proposed class (i.e., that with the lowest degree polynomial) which admits each smoothness level leads us to the sub-class defined by
\begin{align}\label{eq:kernel_class}
K_\alpha(x):=\frac{1}{2(\alpha+1)}\sum_{k=0}^\alpha \frac{|x|^k}{k!}\exp(-|x|), \ \alpha = 1, 2, ...
\end{align}
In the experiments presented in the following section the kernels $K_1$ and $K_4$ will be considered. The kernel $K_1$ is chosen as the simplest differentiable kernel in the class, while $K_4$ is selected as it has efficiency very close to that of the ubiquitous Gaussian kernel. The efficiency of a kernel $K$ relates to the asymptotic mean integrated error which it induces, and may be defined as $\mbox{eff}(K):= (\sigma_K R(K))^{-1}$. It is standard to consider the relative efficiency $\mbox{eff}_{rel}(K) := \mbox{eff}(K)/\mbox{eff}(K^\star)$. The kernel $K^\star$ is the kernel which maximises eff$(K)$ as defined here, and is given by the Epanechnikov kernel. The efficiency and shape of the chosen kernels can be seen in Figure~\ref{fig:Kalpha_plots}, and in relation to the popular Gaussian kernel.

\begin{remark}
The efficiency of a kernel is more frequently defined as the inverse of the definition adopted here. It is considered preferable here to speak of maximising efficiency, rather than minimising it, and hence the above formulation is adopted instead.
\end{remark}

\begin{figure}
\centering
\subfigure[Relative efficiency of $K_\alpha$ for $\alpha = 0, 1, ..., 15$. Relative efficiency of Gaussian kernel (- - - -).]{\includegraphics[width = 8cm]{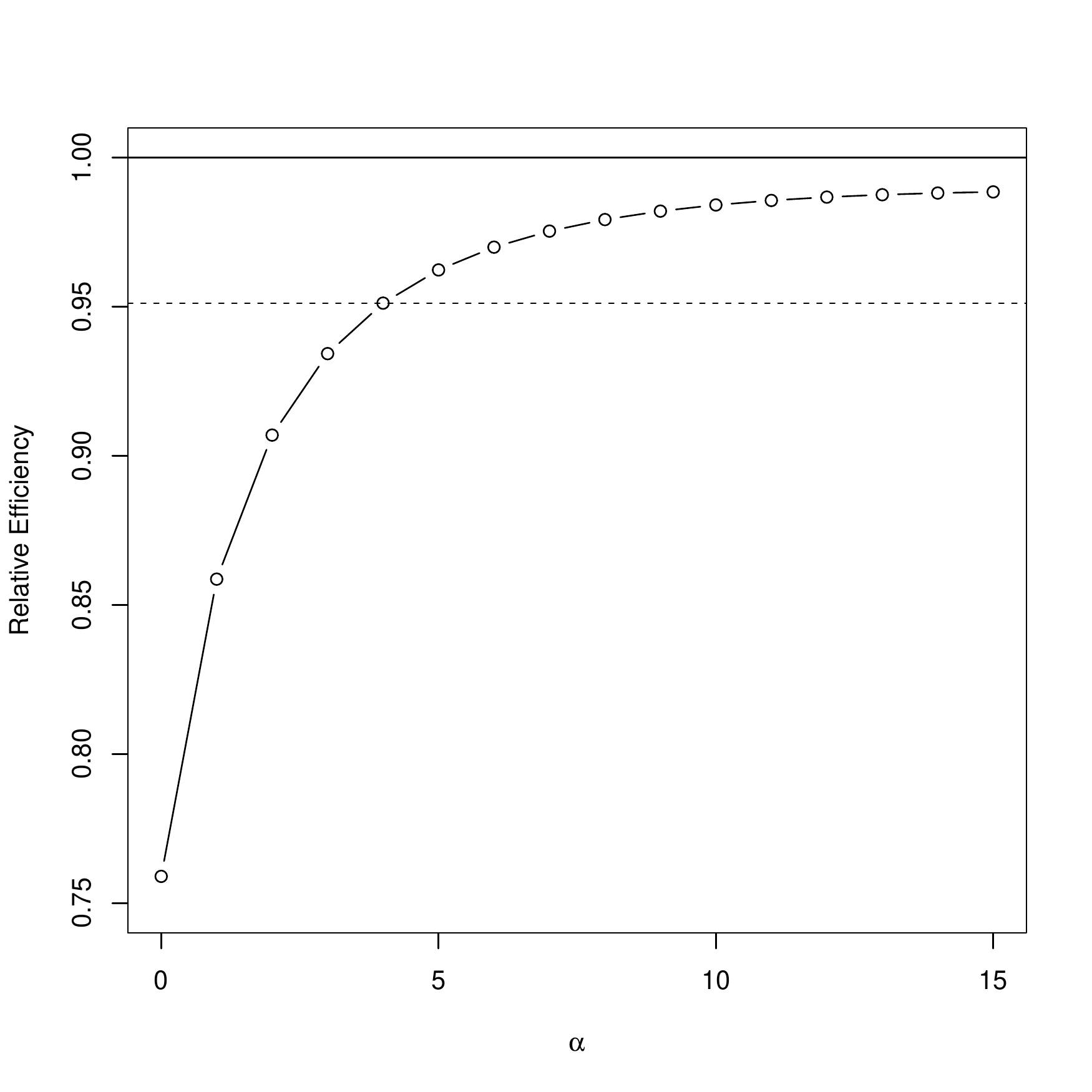}}
\subfigure[Plots of $K_1$ ({\color{blue} - - - -}), $K_4$ ({\color{red}--$\cdot$--$\cdot$--}) and Gaussian~kernel~($\cdots \cdots$)\label{fig:Kalpha_plots}]{\includegraphics[width = 8cm]{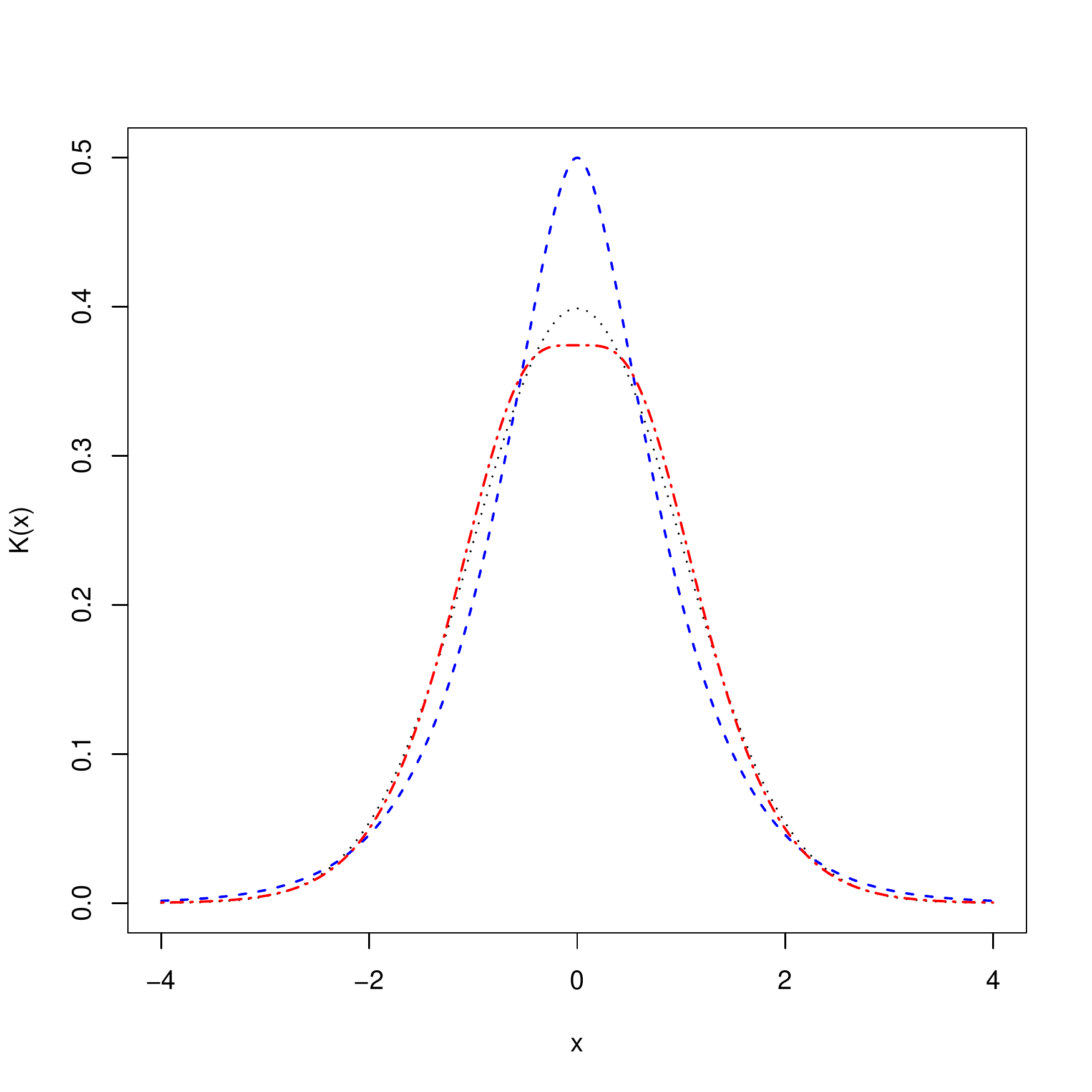}}
\caption{Relative efficiency and shape of the kernels used in experiments \label{fig:Kalpha_plots}}
\end{figure}

\subsection{Density Derivative Estimation}

It is frequently the case that the most important aspects of a density for analysis can be determined using estimates of its derivatives. For example, the roots of the first derivative provide the locations of the stationary points (modes and anti-modes) of the density. In addition pointwise derivatives are useful for determining gradients of numerous projection indices used in projection pursuit~\citep{huber1985projection}. The natural estimate for the $k$-th derivative of $f$ at $x$ is simply,
\begin{align*}
\widehat{f^{(k)}(x)} = \hat f^{(k)}(x) &= \frac{1}{nh}\sum_{i=1}^n \frac{d^k}{dx^k}K\left(\frac{x-x_i}{h}\right)\\
&= \frac{1}{nh^{k+1}}\sum_{i=1}^n K^{(k)}\left(\frac{x-x_i}{h}\right)
\end{align*}
Only the first derivative will be considered explicitly, where higher order derivatives can be simply derived, given an appropriate kernel (i.e., one with sufficiently many derivatives). Considering again the kernels $K_\alpha$ as defined previously, consider that for $\alpha = 1, 2, ...$ we have
\begin{align*}
K_\alpha^{(1)}(x) &= \frac{d}{dx} \frac{1}{2(\alpha+1)}\sum_{k=0}^\alpha \frac{|x|^k}{k!}\exp(-|x|)\\
& = \frac{1}{2(\alpha+1)}\left(\exp(-|x|)\sum_{k=0}^\alpha\frac{k|x|^{k-1}\mbox{sign}(x)}{k!} - \mbox{sign}(x)\exp(-|x|)\sum_{k=0}^\alpha\frac{|x|^k}{k!}\right)\\
& = \frac{\exp(-|x|)}{2(\alpha+1)}\left(\sum_{k=0}^{\alpha-1}\frac{x|x|^{k-1}}{k!} -\sum_{k=0}^\alpha\frac{x|x|^{k-1}}{k!}\right)\\
& = -\frac{\exp(-|x|)}{2(\alpha+1)}\frac{x|x|^{\alpha-1}}{\alpha !} = -\frac{\exp(-|x|)}{2(\alpha+1)!}x|x|^{\alpha-1}\\
\end{align*}
To compute estimates of $\hat f^{(1)}(x)$ only a very slight modification to the methodology discussed previously is required. Specifically, consider that
\begin{align*}
&\sum_{i=1}^n (x-x_i)|x-x_i|^{\alpha-1}\exp\left(-\frac{|x-x_i|}{h}\right)\\ &=\sum_{i=1}^{n(x)}(x-\xo i)^\alpha\exp\left(\frac{\xo i-x}{h}\right) - \sum_{i=n(x)+1}^n (\xo i-x)^\alpha\exp\left(\frac{x-\xo i}{h}\right)\\
&= \sum_{k=0}^\alpha {\alpha \choose k}\left( \exp\left(\frac{\xo {n(x)}-x}{h}\right)x^{\alpha-k}\ell(k, n(x)) - \exp\left(\frac{x-\xo {n(x)}}{h}\right)(- x)^{\alpha-k}r(k, n(x))\right).
\end{align*}
The only difference between this and the corresponding terms in the estimated density is the `-' separating terms in the final expression above. 

\begin{figure}
\centering
\includegraphics[width = 7cm]{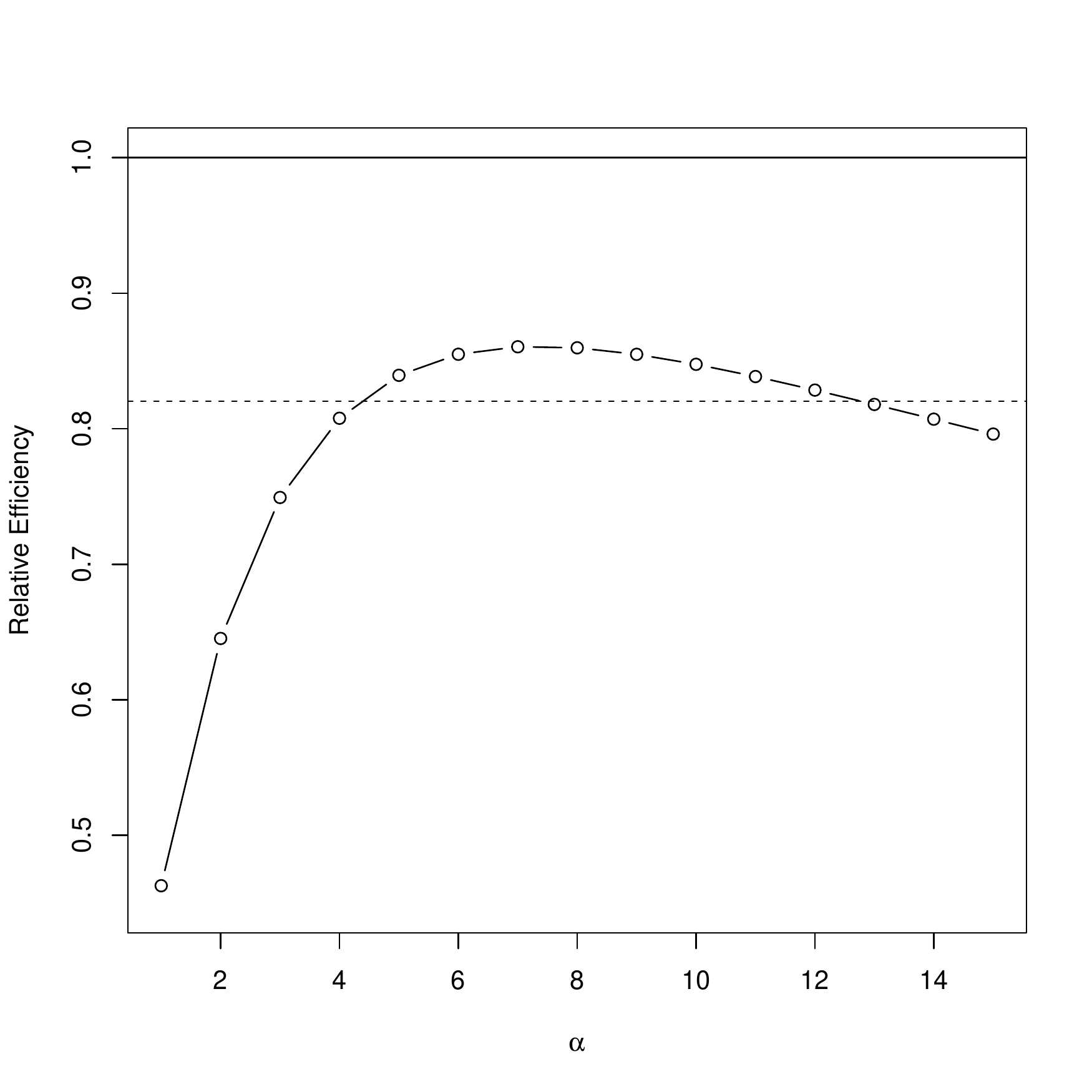}
\caption{Relative efficiency of kernels $K_\alpha$, $\alpha = 1, 2, ..., 15$, for estimating the derivative of a density function. Relative efficiency of Gaussian kernel ( - - - - ) \label{fig:rel_eff_deriv}}
\end{figure}

Now, using the above expression for $K^{(1)}_\alpha(x)$, consider that
\begin{align*}
R(K^{(1)}_\alpha) &= \frac{1}{\left(2(\alpha+1)!\right)^2}\int_{-\infty}^\infty |x|^{2\alpha+2}\exp(-2|x|) dx\\
&= \frac{1}{\left(2(\alpha+1)!\right)^2}\frac{1}{2}\int_{-\infty}^\infty \frac{|x|^{2\alpha+2}}{2^{2\alpha+2}}\exp(-|x|) dx\\
&= \frac{(2\alpha)!}{\left((\alpha+1)!\right)^2}2^{-2\alpha-2}.
\end{align*}
Unlike for the task of density estimation, the relative efficiency of a kernel for estimating the first derivative of a density function is determined in relation to the biweight kernel. The relative efficiency of the adopted class for estimation of the derivative of a density is shown in Figure~\ref{fig:rel_eff_deriv}. The relative efficiency of the Gaussian kernel is again included for context. Again the kernel $K_4$ has similar efficiency to the Gaussian. Here, unlike for density estimation, we can see a clear maximiser with $K_7$. This kernel will therefore also be considered for the task of density derivative estimation in the experiments to follow.

\section{Simulations}\label{sec:sim}

\begin{figure}
\centering
\subfigure[Gaussian]{\includegraphics[width = 3.85cm]{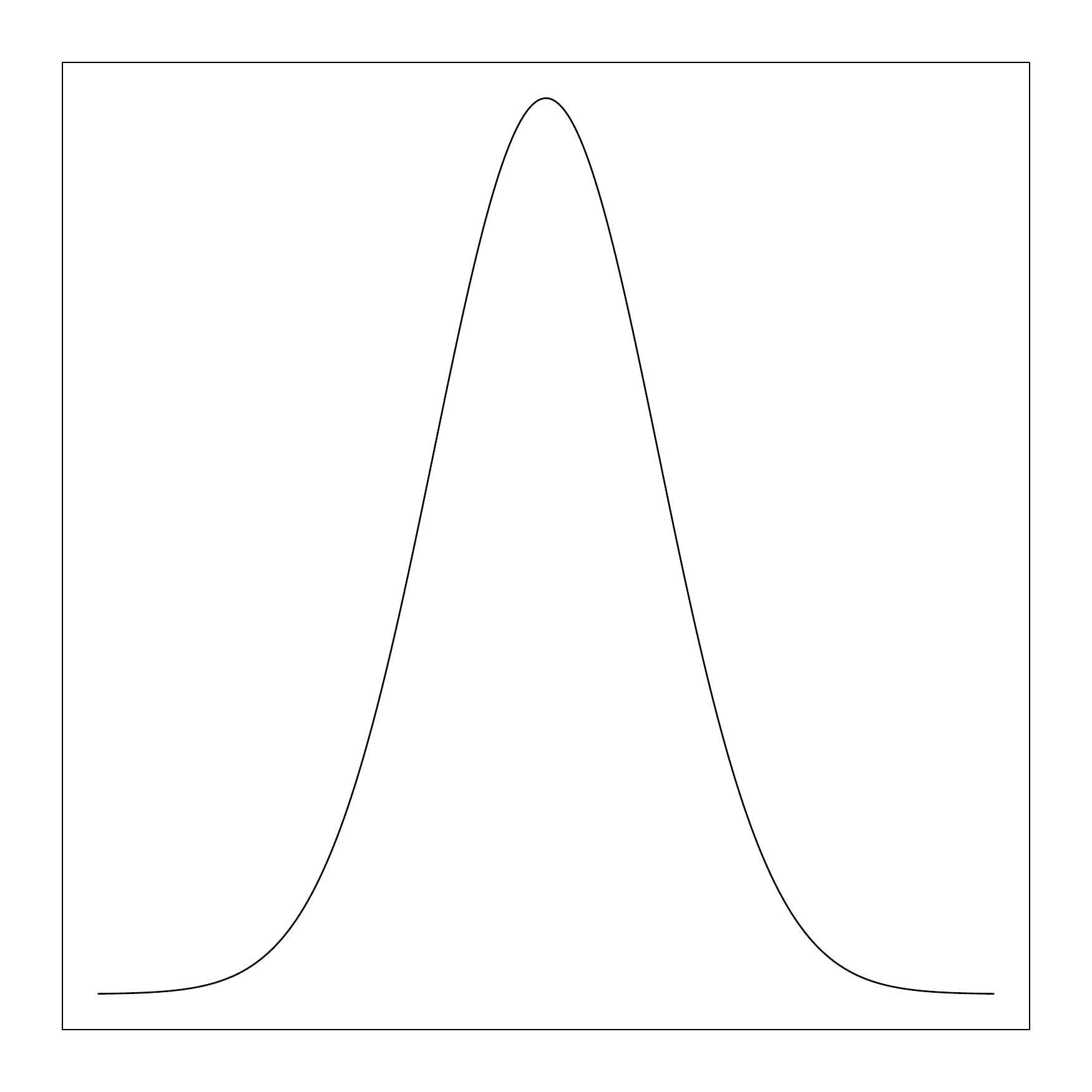}}
\subfigure[Uniform]{\includegraphics[width = 3.85cm]{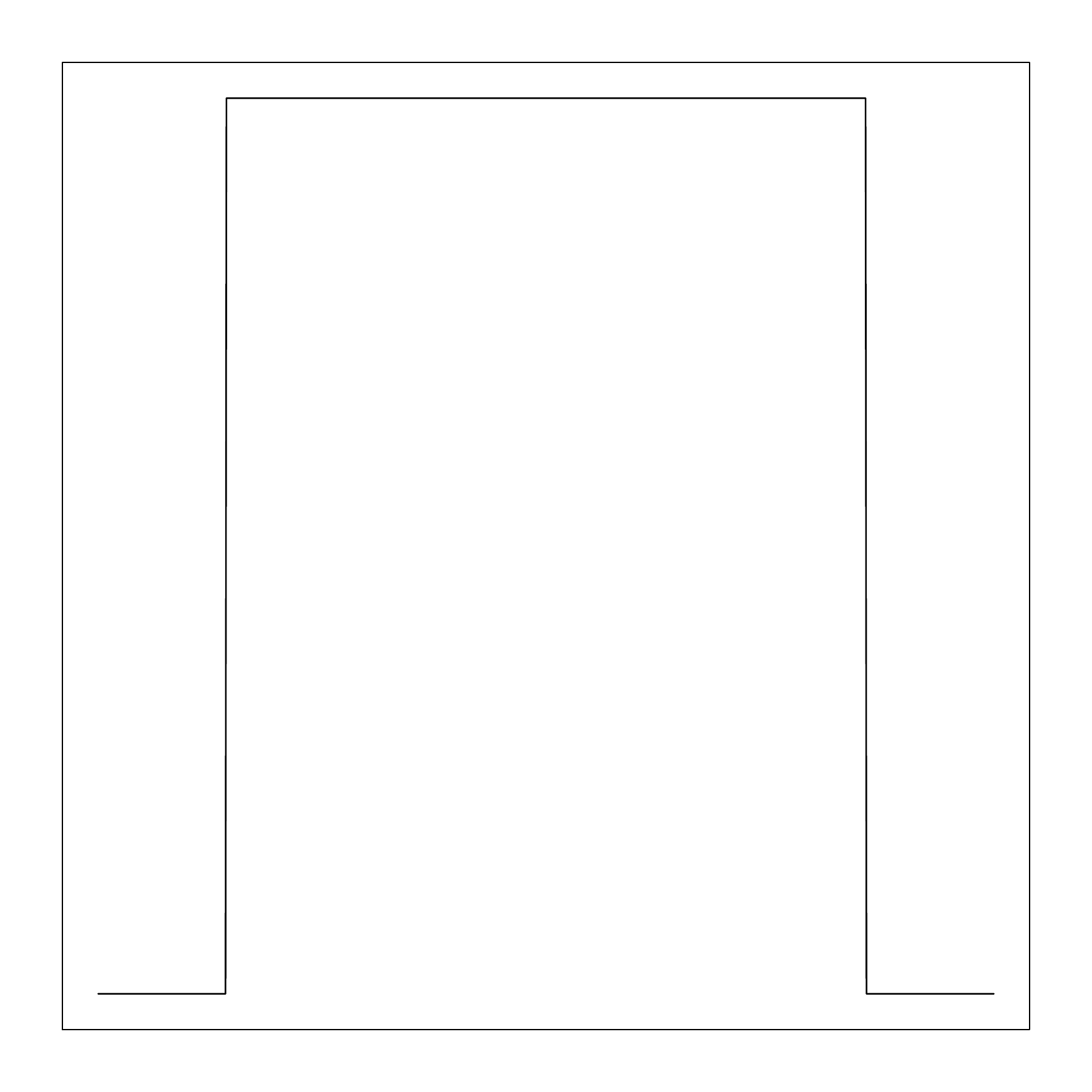}}
\subfigure[Scale mixture]{\includegraphics[width = 3.85cm]{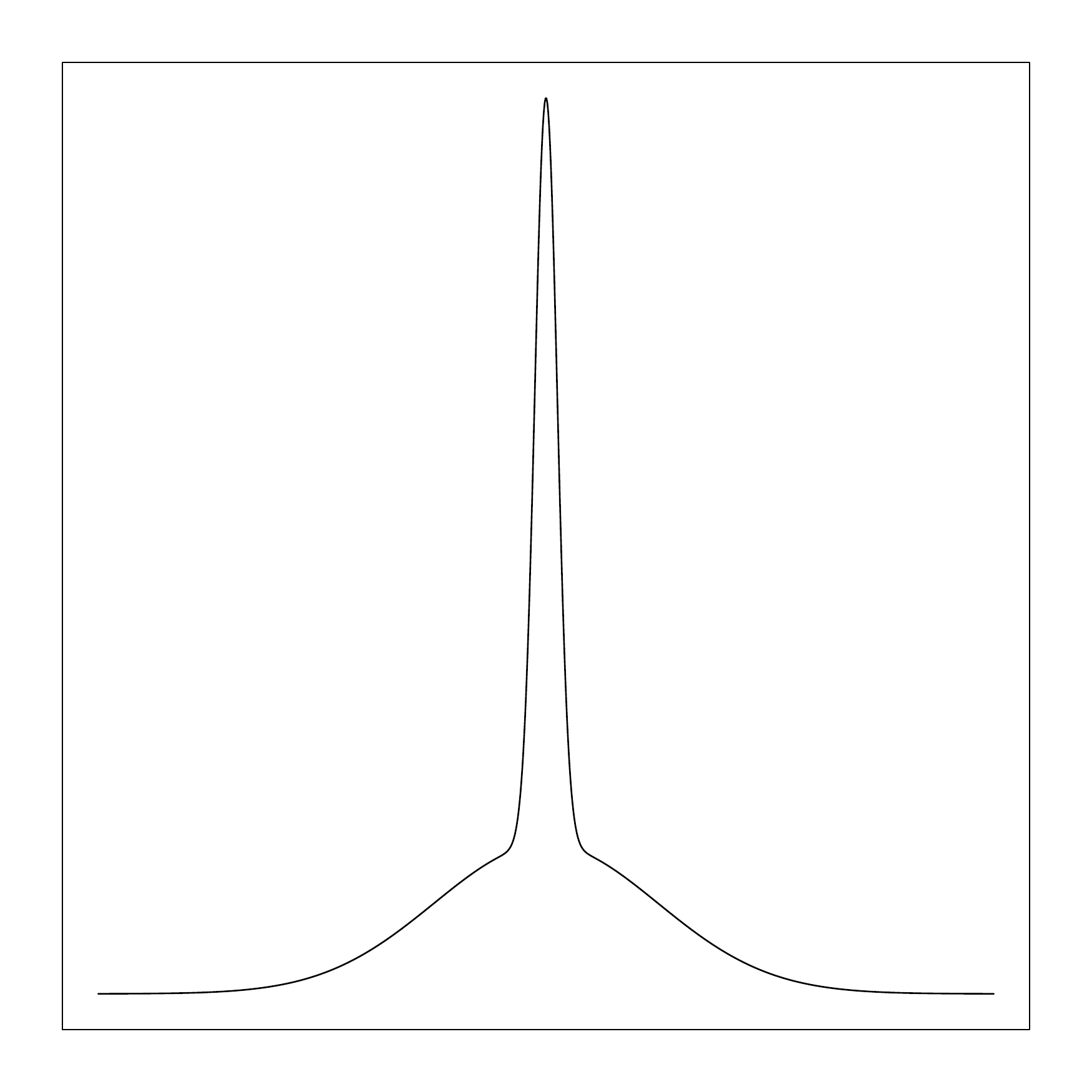}}
\subfigure[Simple bimodal]{\includegraphics[width = 3.85cm]{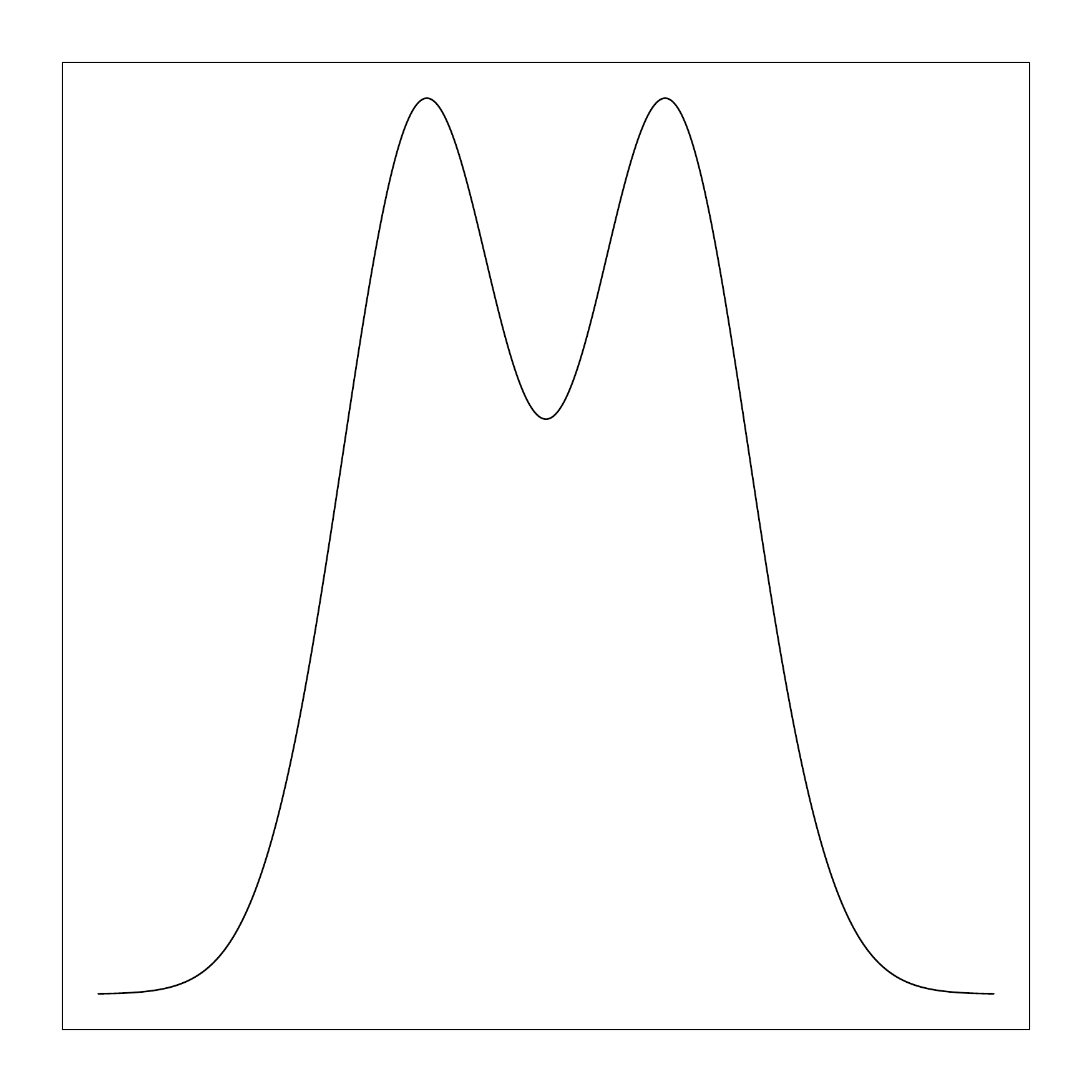}}
\subfigure[Skew]{\includegraphics[width = 3.85cm]{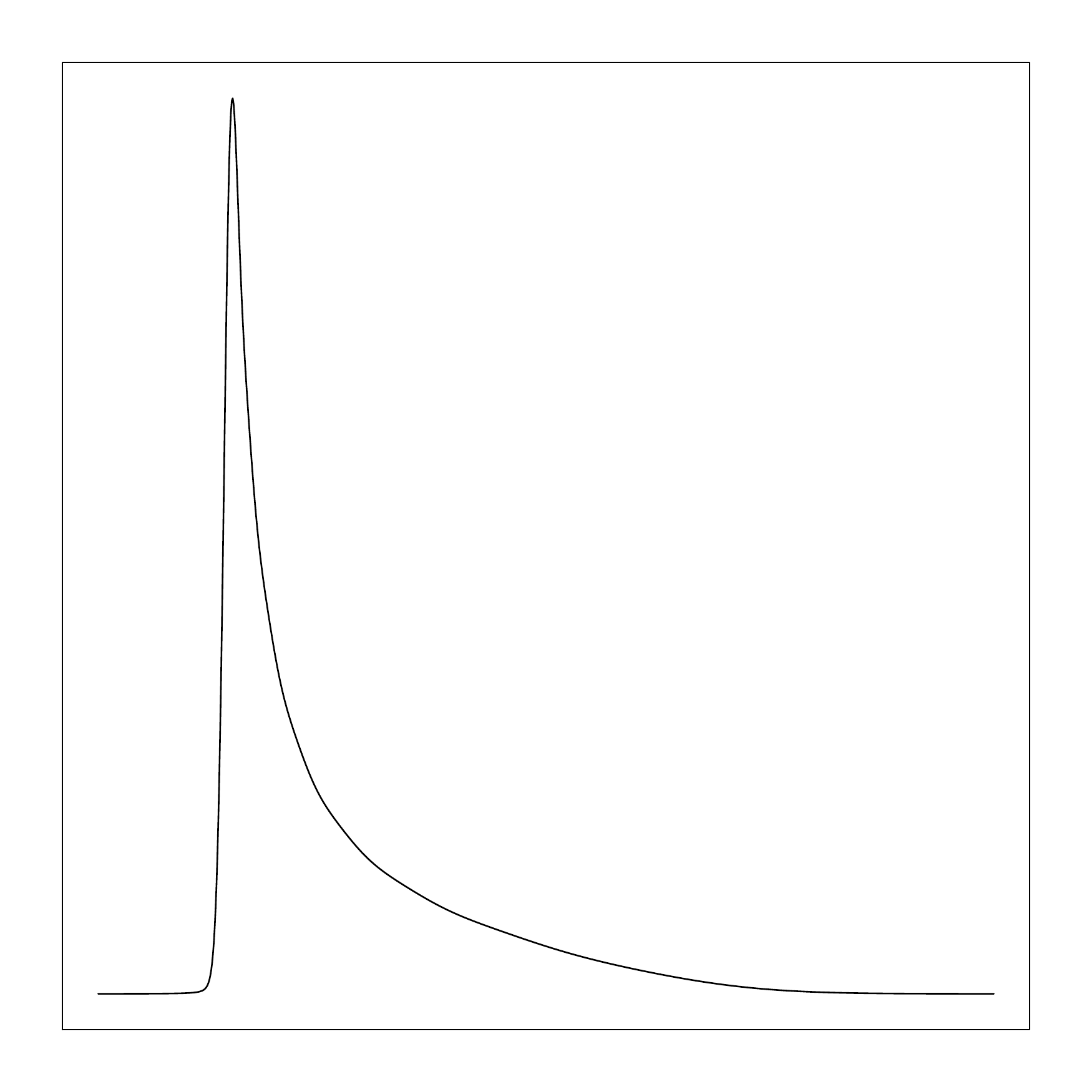}}
\subfigure[Spiked bimodal]{\includegraphics[width = 3.85cm]{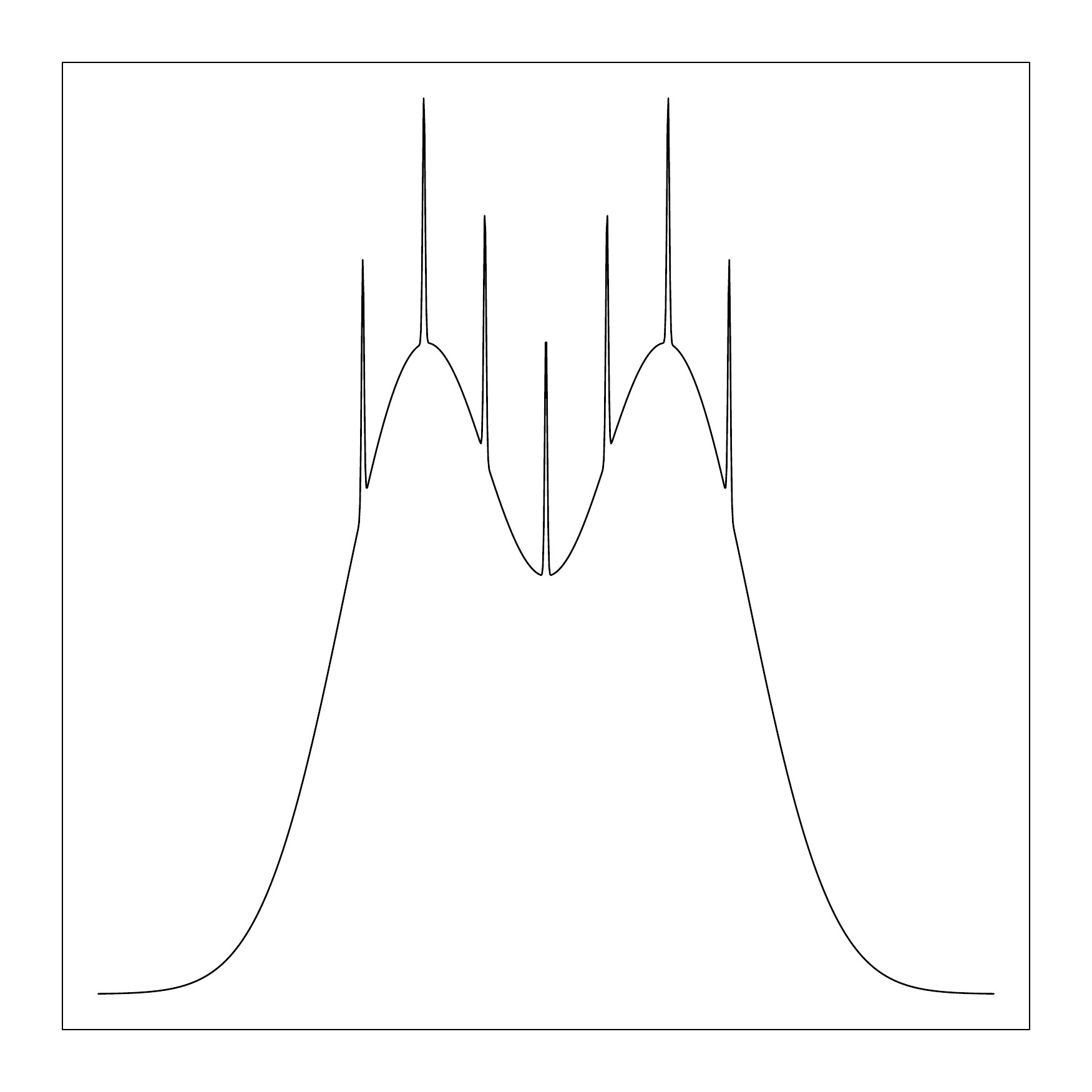}}
\subfigure[Claw]{\includegraphics[width = 3.85cm]{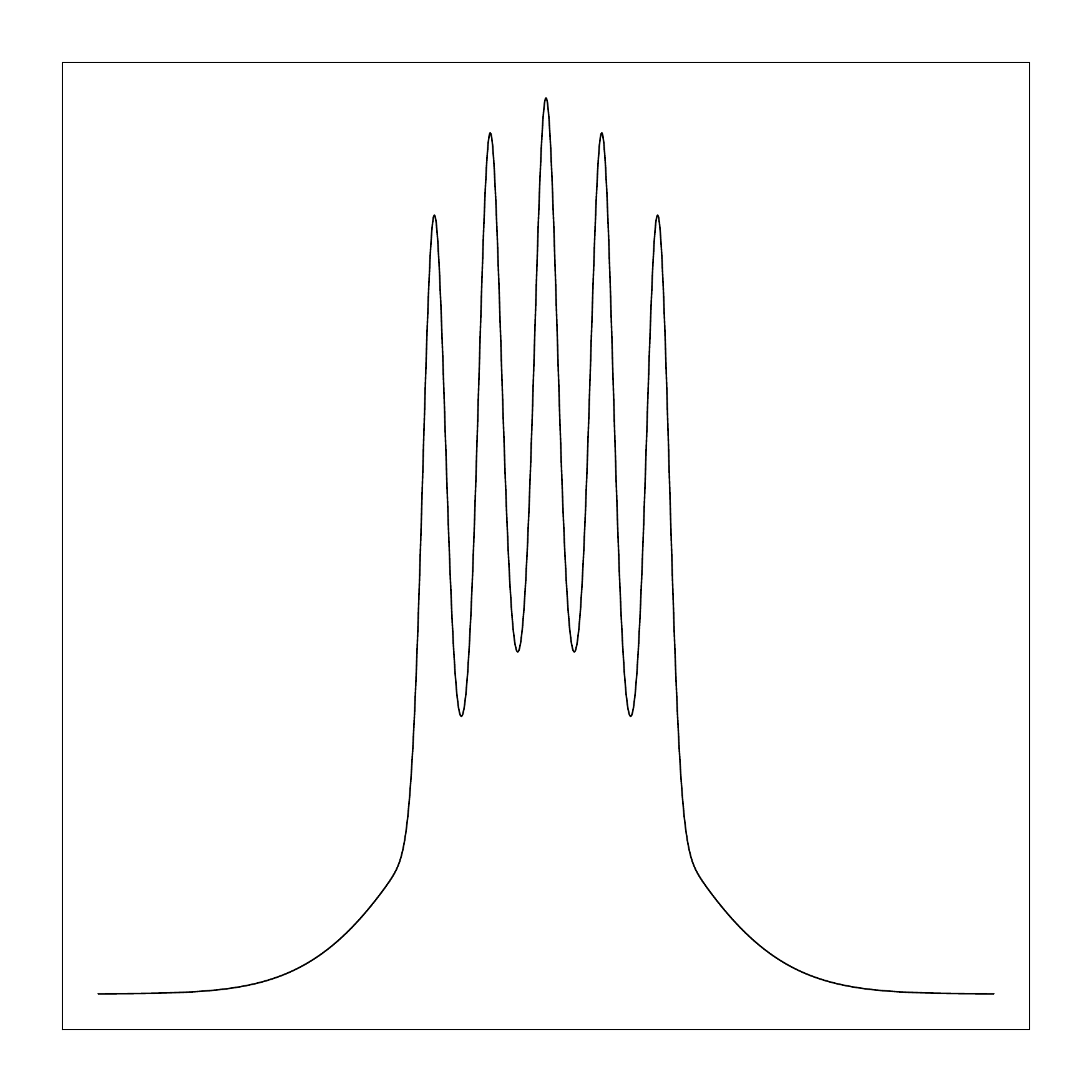}}
\subfigure[Skew bimodal]{\includegraphics[width = 3.85cm]{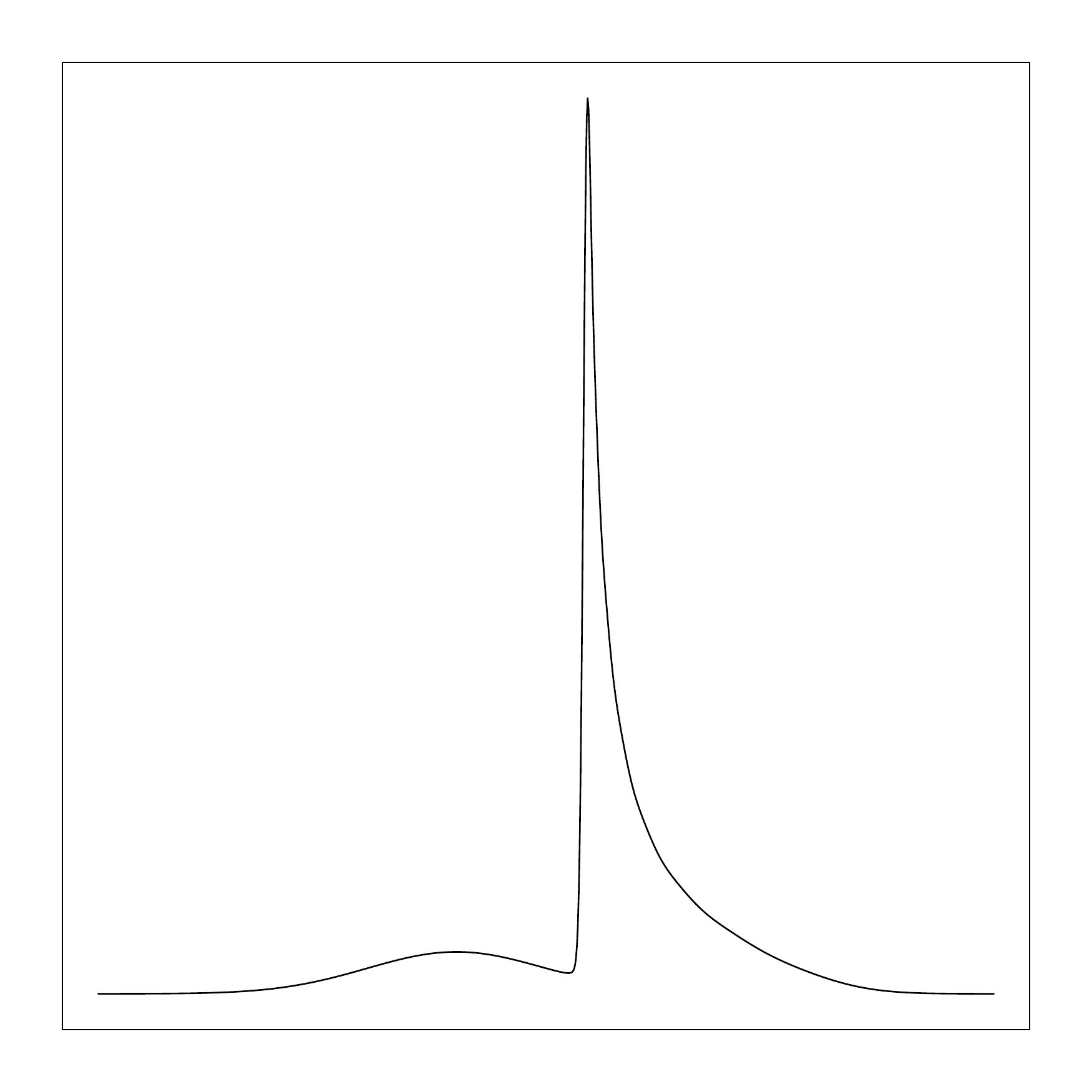}}
\caption{Collection of densities used in experiments.\label{fig:densities}}
\end{figure}

This section presents the results from a thorough simulation study conducted to illustrate the efficiency and effectiveness of the proposed approach for density and density derivative estimation. 
A collection of eight univariate densities are considered, many of which are taken from the popular collection of benchmark densities given in~\cite{MarronW1992}. Plots of all eight densities are given in Figure~\ref{fig:densities}

For context, comparisons will be made with the following existing methods. For these the Gaussian kernel was used, as the most popular kernel used in the literature.
\begin{enumerate}
\item The exact estimator using the Gaussian kernel, for which the implementation in the {\tt R} package {\tt kedd}~\citep{keddpkg} was used. This approach was only applied to samples with fewer than 10~000 observations, due to the high computation time required for large samples.
\item The binned estimator with Gaussian kernel using the package {\tt KernSmooth}~\citep{KernSmoothpkg}.
\item The fast Fourier transform using {\tt R}'s base {\tt stats} package.
\item The truncated Taylor expansion approach~\citep{RaykarDZ2010}, for which a wrapper was created to implement the authors' {\tt c++} code\footnote{the authors' code was obtained from \url{https://www.umiacs.umd.edu/user.php?path=vikas/Software/optimal_bw/optimal_bw_code.htm}} from within {\tt R}.
\end{enumerate}
The main computational components of the proposed method were implemented in {\tt c++}, with the master functions in {\tt R} via the {\tt Rcpp}~\citep{Rcpppkg} package\footnote{A simple {\tt R} package is available from \url{https://github.com/DavidHofmeyr/fkde}}. The binned estimator using the proposed class of kernels will also be considered. Because of the nature of the kernels used, as discussed in Section~\ref{sec:kernels}, the computational complexity of the corresponding binned estimator is $\mathcal{O}(n + (\alpha+1)b)$, where $b$ is the number of bins.

Accuracy will be assessed using the integrated squared error between the kernel estimates and the true sampling densities (or their derivatives), i.e., $||\hat f^{(k)} - f^{(k)}||^2_2 = \int_{-\infty}^\infty(\hat f^{(k)}(x) - f(x)^{(k)})^2 dx$. 
Exact evaluation of these integrals is only possible for very specific cases, and so they are numerically integrated.
For simplicity in all cases Silverman's rule of thumb is used to select the bandwidth parameter~\citep{silverman2018density}. This extremely popular heuristic is motivated by the optimal asymptotic mean integrated squared error (AMISE) bandwidth value. The heuristic is most commonly applied to density estimation, where the direct extension to the first two derivatives will also be used herein. For kernel $K$ the AMISE optimal bandwidth is given by
\begin{align*}
\hamise = \left(\frac{(2k+1)R(K^{(k)})}{\sigma_K^4 R(f^{(k+2)})n}\right)^{1/(2k+5)}.
\end{align*}
This objective is generally preferred over the mean integrated squared error as it reduces the dependency on the underlying unknown density function to only the functional $R(f^{(k+2)})$. Silverman's heuristic replaces $R(f^{(k+2)})$ with $R(\phi_{\hat\sigma}^{(k+2)})$, where $\phi_\sigma$ is the normal density with scale parameter $\sigma$. The scale estimate $\hat\sigma$ is computed from the observations, usually as their standard deviation.

\subsection{Density Estimation}

In this subsection the accuracy and efficiency of the proposed method for density estimation are investigated. 

\subsubsection{Evaluation on a Grid}

Many of the approximation methods for kernel density estimation necessitate that the evaluation points, $\{\tilde x_1, ..., \tilde x_m\}$, are equally spaced~\citep{ScottS1985,Silverman1982}. In addition such an arrangement is most suitable for visualisation purposes. Here the speed and accuracy of the various methods for evaluation/approximation of the density estimates are considered, where evaluation points are restricted to being on a grid. 

\paragraph{Accuracy:} The accuracy of all methods is reported in Table~\ref{tb:ISE-nvar}. Sixty samples were drawn from each density, thirty of size 1~000 and thirty of size 1~000~000. The number of evaluation points was kept fixed at 1~000. The estimated mean integrated squared error is reported in the table. The lowest average is highlighted in each case. The error values for all methods utilising the Gaussian kernel ($\phi)$ are extremely similar, which attests to the accuracy of the approximate methods. The error of kernel $K_4$ is also very similar to that of the Gaussian kernel methods. This is unsurprising due to its similar efficiency value. The kernel $K_1$ obtains the lowest error over all and in the most cases.

\begin{table}
\caption{Estimated mean integrated squared error of density estimates from 30 replications. Sets of 1~000 and 1~000~000 observations are considered. Lowest average error for each scenario is highlighted in bold. Apparent ties were broken by considering more significant figures.\label{tb:ISE-nvar}}
\begin{center}
\scalebox{.75}{
\begin{tabular}{ll|cccccccccccc}
&&\multicolumn{8}{c}{Method}\\
\cline{3-10}
 & Density &  Exact $\phi$ &  Tr. Taylor $\phi$ &  Binned $\phi$ &  FFT $\phi$ &  Exact $K_1$ &  Exact $K_4$ &  Binned $K_1$ &  Binned $K_4$ \\
\hline
(a) & n=1e+03  & 9.63e-04 & 9.66e-04 & 9.62e-04 & {\bf 9.55e-04} & 1.08e-03 & 9.62e-04 & 1.07e-03 & 9.62e-04 \\
 & n=1e+06 & -  & 5.1e-06 & 5.1e-06 & {\bf 4.73e-06} & 5.54e-06 & 5.1e-06 & 5.54e-06 & 5.1e-06 \\
(b) & n=1e+03  & 3.81e-02 & {\bf 2.97e-02} & 3.81e-02 & 3.81e-02 & 3.44e-02 & 3.88e-02 & 3.44e-02 & 3.88e-02 \\
 & n=1e+06 & -  & {\bf 6.85e-03} & 9.02e-03 & 9.02e-03 & 8e-03 & 9.18e-03 & 8e-03 & 9.18e-03 \\
(c) & n=1e+03  & 1.02e-01 & 1.02e-01 & 1.02e-01 & 1.02e-01 & {\bf 8.33e-02} & 1.05e-01 & 8.33e-02 & 1.05e-01 \\
 & n=1e+06 & -  & 3.73e-03 & 3.76e-03 & 4.01e-03 & {\bf 3.54e-03} & 3.72e-03 & 3.57e-03 & 3.75e-03 \\
(d) & n=1e+03  & 1.8e-03 & 1.8e-03 & 1.8e-03 & 1.8e-03 & {\bf 1.79e-03} & 1.79e-03 & 1.79e-03 & 1.79e-03 \\
 & n=1e+06 & -  & 1.05e-05 & 1.05e-05 & {\bf 1e-05} & 1.13e-05 & 1.05e-05 & 1.14e-05 & 1.05e-05 \\
(e) & n=1e+03  & 1.02e-01 & 9.13e-02 & 1.02e-01 & 1.02e-01 & 8.68e-02 & 1.05e-01 & {\bf 8.68e-02} & 1.05e-01 \\
 & n=1e+06 & -  & 9.39e-03 & 9.42e-03 & 9.69e-03 & {\bf 8.07e-03} & 9.4e-03 & 8.11e-03 & 9.44e-03 \\
(f) & n=1e+03  & 3.54e-03 & 3.54e-03 & 3.54e-03 & 3.53e-03 & 3.51e-03 & 3.54e-03 & {\bf 3.51e-03} & 3.54e-03 \\
 & n=1e+06 & -  & 1.26e-03 & 1.26e-03 & 1.26e-03 & {\bf 1.17e-03} & 1.29e-03 & 1.17e-03 & 1.29e-03 \\
(g) & n=1e+03  & 4.3e-02 & 4.3e-02 & 4.3e-02 & 4.3e-02 & {\bf 3.62e-02} & 4.51e-02 & 3.62e-02 & 4.51e-02 \\
 & n=1e+06 & -  & 2.34e-03 & 2.36e-03 & 2.49e-03 & {\bf 2.22e-03} & 2.33e-03 & 2.23e-03 & 2.35e-03 \\
(h) & n=1e+03  & 1.02e-01 & 1.02e-01 & 1.02e-01 & 1.02e-01 & {\bf 8.87e-02} & 1.04e-01 & 8.87e-02 & 1.04e-01 \\
 & n=1e+06 & -  & 1.77e-02 & 1.77e-02 & 1.82e-02 & {\bf 1.47e-02} & 1.79e-02 & 1.48e-02 & 1.79e-02 \\
 \hline
\end{tabular}
}
\end{center}
\end{table}

\begin{figure}
\centering
\subfigure[$n = 1000$]{\includegraphics[width = 6.5cm]{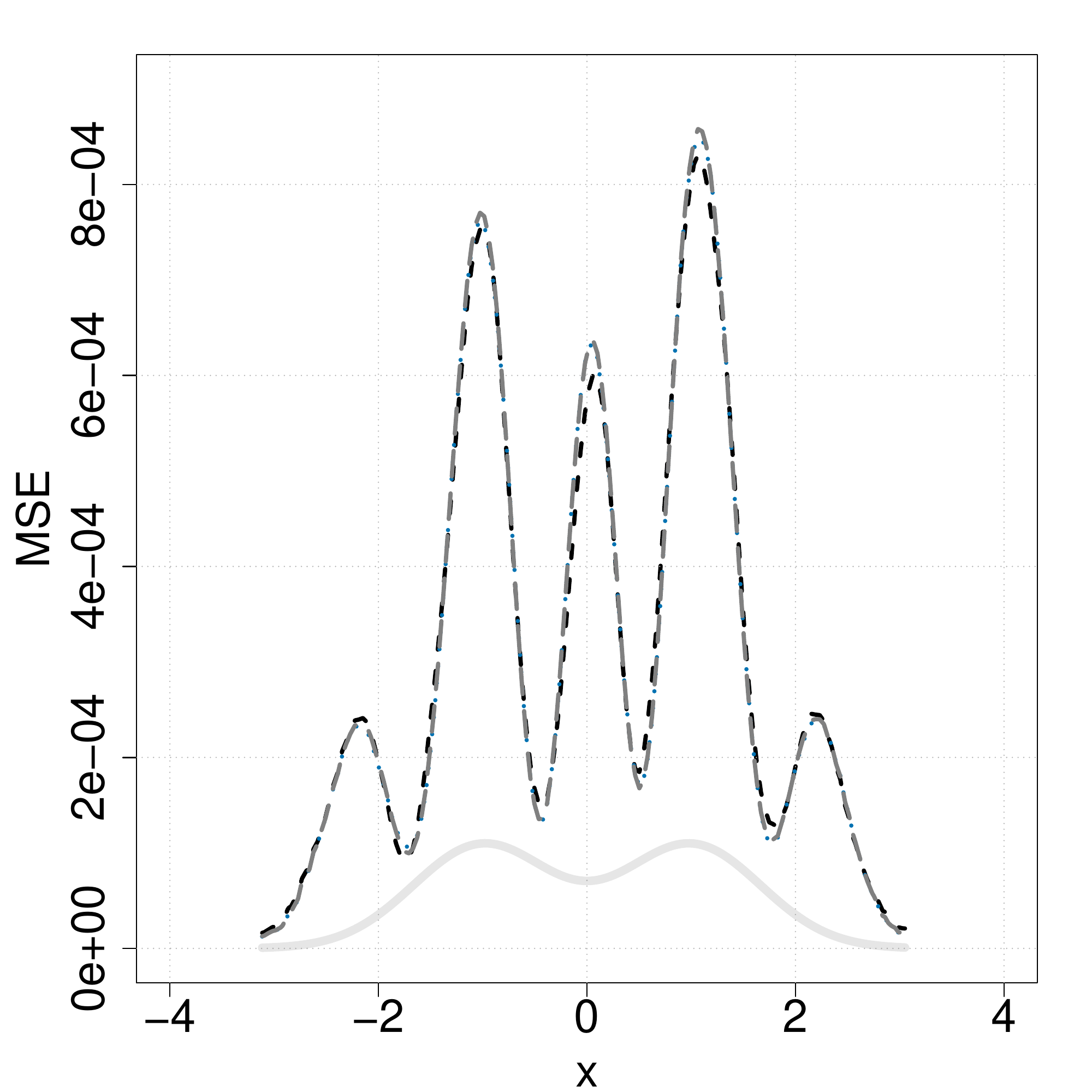}\label{fig:denDMSE1000}}
\subfigure[$n = 1000000$]{\includegraphics[width = 6.5cm]{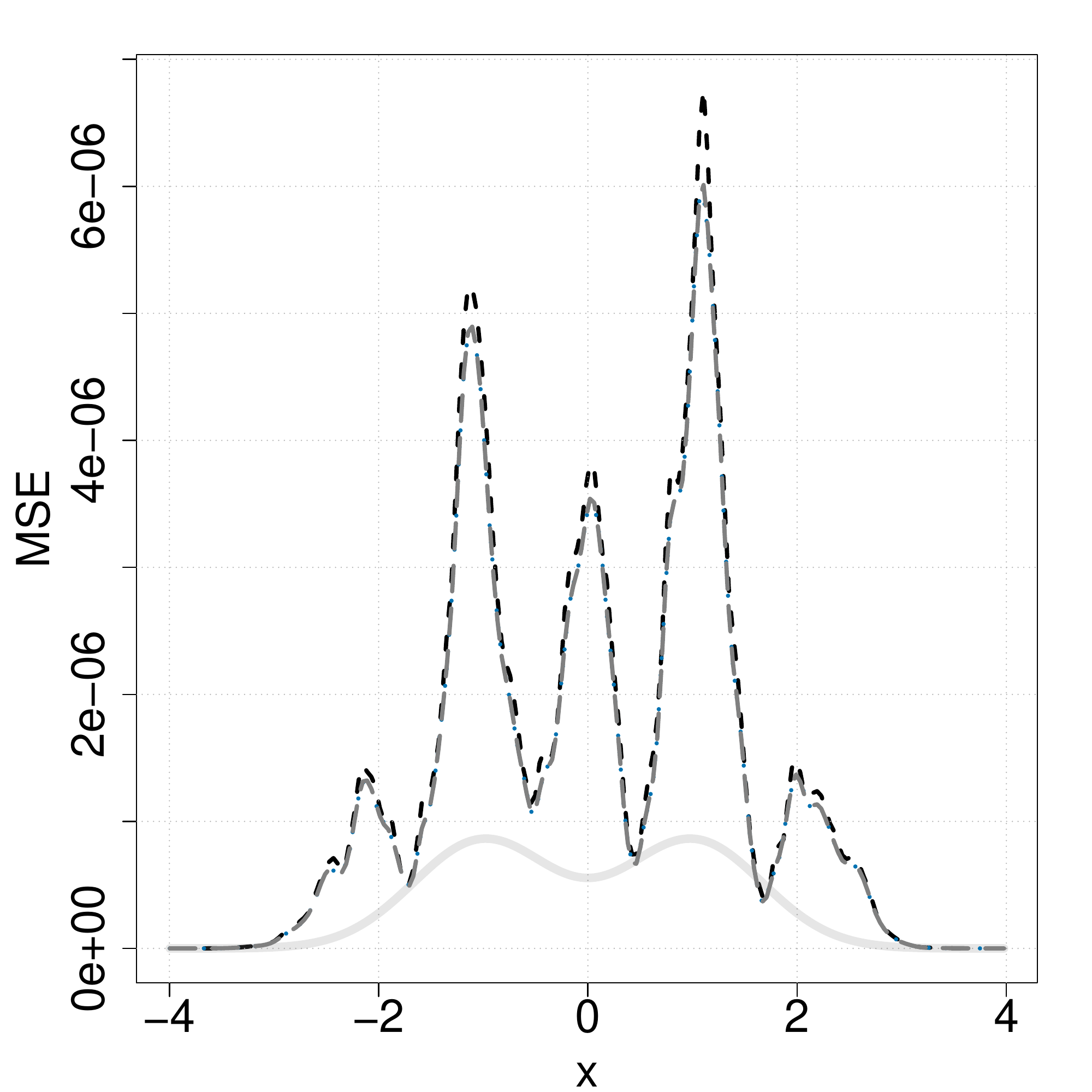}\label{fig:denDMSE1000000}}\\
\small{Tr. Taylor $\phi$ ({\color{col3} --- ---}), Exact $K_1$ ({\color{col2} -- -- --}), Exact~$K_4$~({\color{col1} $\cdots \cdots$})}

\caption{Estimated pointwise mean squared error for density (d).\label{fig:denDMSE}}
\end{figure}

In addition the estimated pointwise mean squared error for density (d) was computed for the exact estimates using kernels $K_1$ and $K_4$ and the truncated Taylor approximation for the Gaussian kernel estimate. These can be seen in Figure~\ref{fig:denDMSE}. In addition the shape of density (d) is shown. This density was chosen as it illustrates the improved relative performance of more efficient kernels as the sample size increases. For the smaller sample size kernel $K_1$ has a lower estimated mean integrated squared error, which is evident in Figure~\ref{fig:denDMSE1000}. The mean squared error for the other two methods is almost indistinguishable. On the other hand for the large sample size, shown in Figure~\ref{fig:denDMSE1000000}, the error for kernel $K_1$ is noticeably larger at the extrema of the underlying density than $K_4$ and the Gaussian approximation. A brief discussion will be given in the discussion to follow in relation to kernel efficiency and the choice of kernel.

\begin{figure}
\centering
\subfigure[Fixed number of evaluation points, increasing sample size]{\includegraphics[width = 8cm]{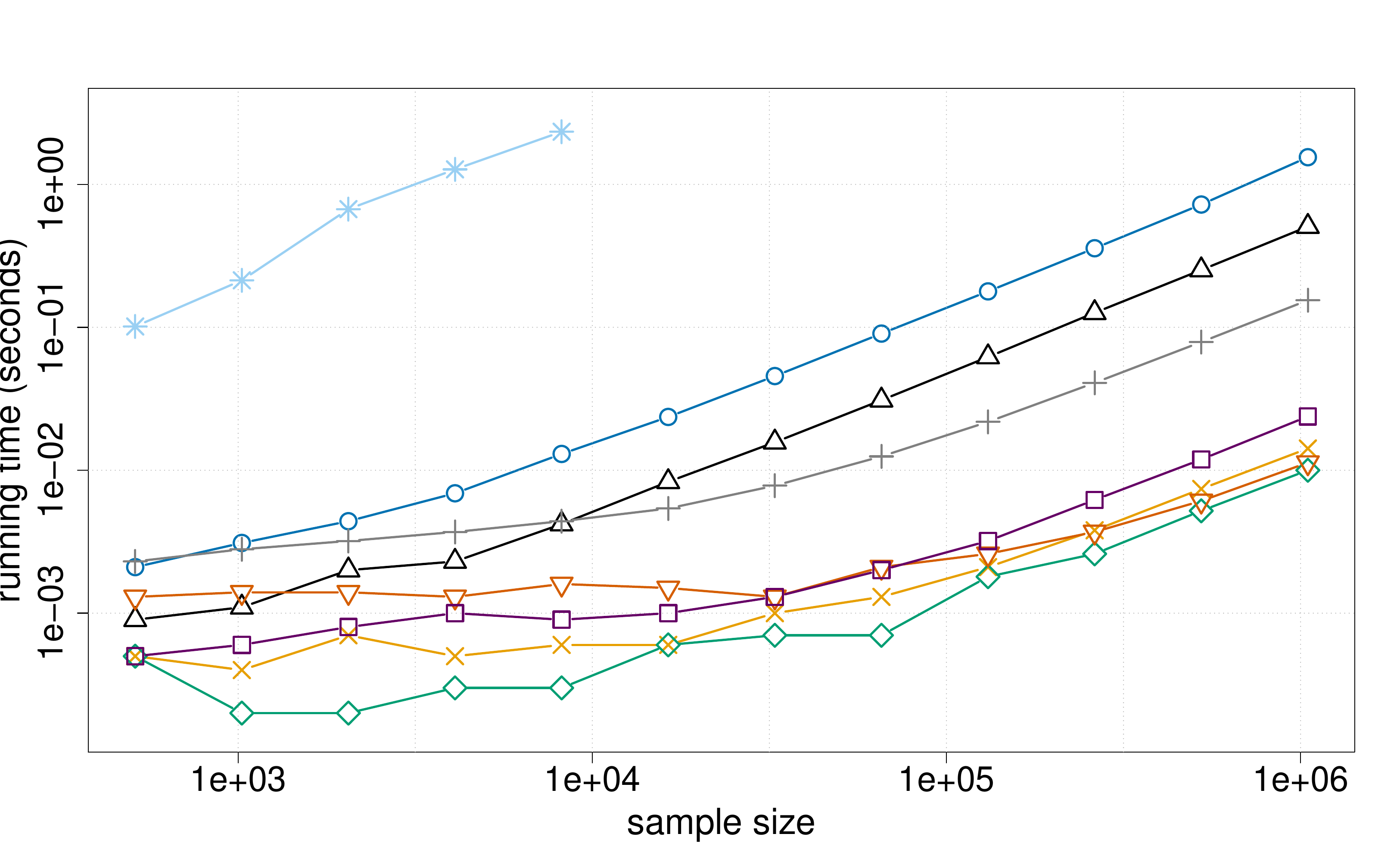}\label{fig:runtime_nvar}}
\subfigure[Fixed number of observations, increasing number of evaluation points]{\includegraphics[width = 8cm]{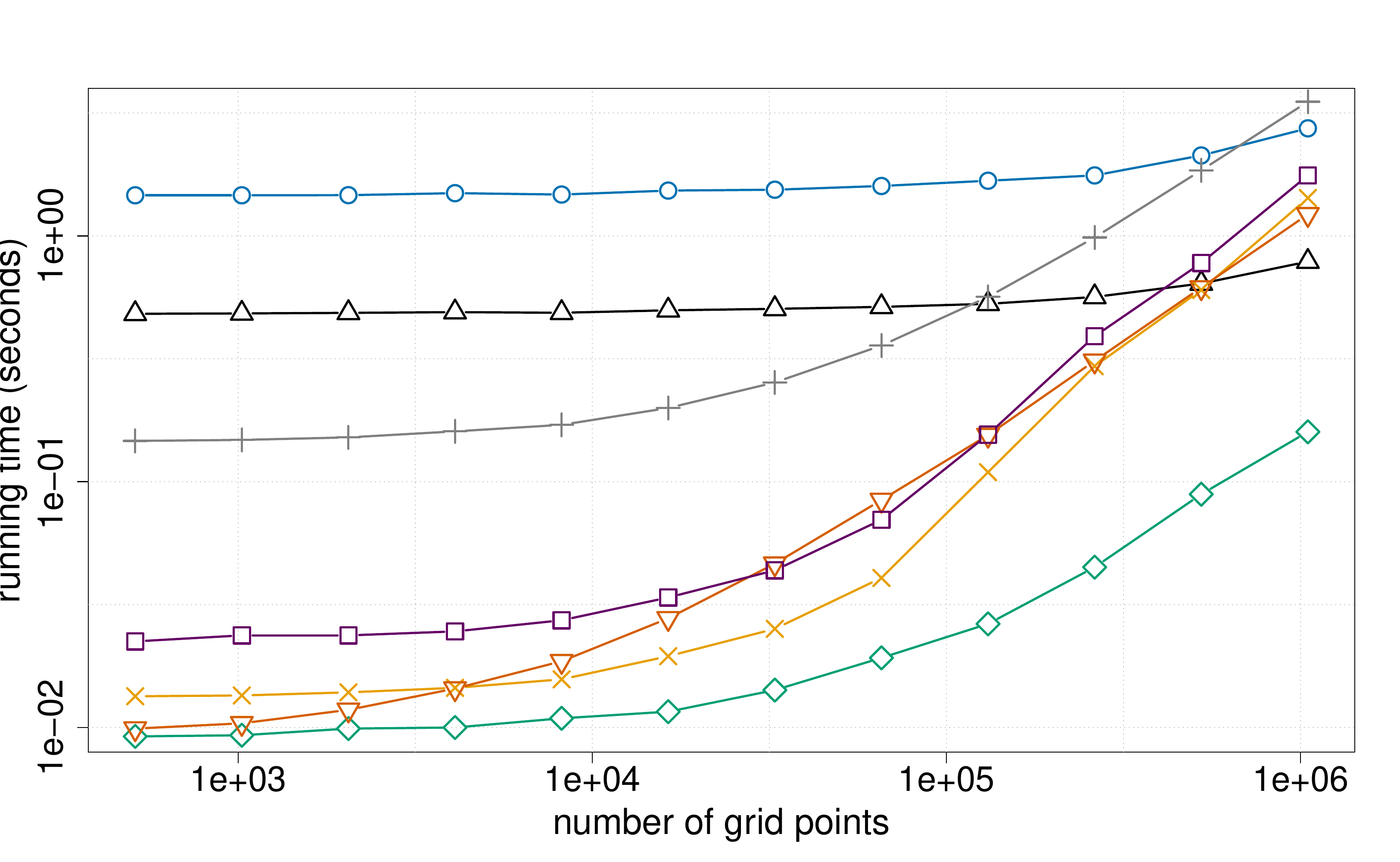}\label{fig:runtime_gridvar}}
\vspace{10pt}
\small{Exact $\phi$ ({\color{col8} --$\ast$--}), Tr. Taylor $\phi$ ({\color{col3} --$+$--}), Binned $\phi$ ({\color{col4} --$\times$--}), FFT $\phi$ ({\color{col7} --$\square$--}), Exact $K_1$ ({\color{col2} --$\triangle$--}), Exact~$K_4$~({\color{col1} --$\circ$--}), Binned $K_1$ ({\color{col5} --$\diamond$--}), Binned $K_4$ ({\color{col9} --$\triangledown$--})}

\caption{Computation times for density (d) evaluated on a grid\label{fig:runtime}}
\end{figure}

\paragraph{Computational efficiency:} The running times for all densities are extremely similar, and more importantly the comparative running times between different methods are almost exactly the same accross the different densities. It is therefore sufficient for comparisons to consider a single density. Note that in order to evaluate the density estimate at a point not in the sample, the proposed approach requires all computations needed to evaluate the density at the sample points. Evaluation on a grid may therefore be seen as something of a worst case for the proposed approach. However, once the decision to evaluate the density estimate at points other than the sample points has been made, the marginal cost of increasing the number of evaluation points is extremely small. This fact is well captured by Figure~\ref{fig:runtime}. This figure shows plots of the average running times from the methods considered when applied to density (d), plotted on a log-scale. Figure~\ref{fig:runtime_nvar} shows the effect of increasing the number of observations, while keeping the number of evaluation points fixed at 1~000. On the other hand Figure~\ref{fig:runtime_gridvar} shows the case where the number of observations is kept fixed (at 100~000) and the number of evaluation points is increased. In the former the proposed method, despite obtaining an exact evalution of the estimate density, is reasonably competitive with the slower of the approximate methods. It is also orders of magnitude faster than the exact method using the Gaussian kernel. In the latter it can be seen that as the number of evaluation points increases the proposed exact approach is even competitive with the fastest approximate methods.

Overall the binned approximations provide the fastest evaluation. The nature of the proposed kernels and the proposed method for fast evaluation means that the corresponding binned estimators (particularly that pertaining to kernel $K_1$) are extremely computationally efficient.

\subsubsection{Evaluation at the Sample Points}

Evaluation of the estimated density at the sample points themselves has important applications in, among other things, computation of non-parametric pseudo-likelihoods and in the estimation of sample entropy. Of the methods considered only the exact methods and the truncated Taylor expansion approximation are applicable to this problem. Table~\ref{tb:ISE-n-smp-BWBS} shows the average integrated squared error of the estimated densities from 30 replications for each sampling scenario. Unsuprisingly the accuracy values and associated conclusions are similar to those for the grid evaluations above. An important difference is that when the density estimates are required at all of the sample values, the proposed exact method outperforms the approximate method in terms of computation time. This is seen in Table~\ref{tb:runtime-n-smp-BWBS}, where the average running times for all densities are reported. The exact evalution for kernel $K_4$ is similar to the truncated Taylor approximate method, while the exact evaluation using kernel $K_1$ is roughly five times faster with the current implementations.

\begin{remark}
It is important to reiterate the fact that the proposed approach is exact. This exactness becomes increasingly important when these density estimates form part of a larger routine, such as maximum pseudo-likelihood or in projection pursuit. When the density estimates are only approximate it becomes more difficult to determine how changes in the sample points, or in hyperparameters, will affect these estimated values.
\end{remark}

\begin{table}
\caption{Average integrated squared error of density estimates from 30 replications. Sets of 1~000 and 100~000 observations are considered. Evaluation is conducted for the entire collection of sample points in each case. Lowest average error for each scenario is highlighted in bold. Apparent ties were broken by considering more significant figures.\label{tb:ISE-n-smp-BWBS}}
\begin{center}
\scalebox{.7}{
\begin{tabular}{ll|cccccccccccc}
&&\multicolumn{8}{c}{Density}\\
\cline{3-10}
&& (a) & (b) & (c) & (d) & (e) & (f) & (g) & (h)\\ 
\hline
Exact Gauss & n = 1e+03 & 8.29e-04 & 2.18e-02 & 1.03e-01 & 1.86e-03 & 1.02e-01 & 3.43e-03 & 4.32e-02 & 1.03e-01\\
Trunc. Taylor Gauss & n = 1e+03 & 8.29e-04 & 2.18e-02 & 1.03e-01 & 1.86e-03 & 1.02e-01 & 3.43e-03 & 4.32e-02 & 1.03e-01\\
 & n = 1e+05 & 2.75e-05 & 7.23e-03 & 1.58e-02 & 6.38e-05 & 2.64e-02 & 1.45e-03 & 9.81e-03 & 3.78e-02\\
Exact $K_1$ & n = 1e+03 & 9.17e-04 & {\bf 2.01e-02} & {\bf 8.53e-02} & {\bf 1.84e-03} & {\bf 8.76e-02} & {\bf 3.41e-03} & {\bf 3.64e-02} & {\bf 9.03e-02}\\
 & n = 1e+05 & 3.03e-05 & {\bf 6.44e-03} & {\bf 1.37e-02} & 6.79e-05 & {\bf 2.2e-02} & {\bf 1.39e-03} & {\bf 8.36e-03} & {\bf 3.16e-02}\\
Exact $K_4$ & n = 1e+03 & {\bf 8.28e-04} & 2.21e-02 & 1.06e-01 & 1.85e-03 & 1.04e-01 & 3.43e-03 & 4.53e-02 & 1.06e-01\\
 & n = 1e+05 & {\bf 2.75e-05} & 7.36e-03 & 1.57e-02 & {\bf 6.37e-05} & 2.68e-02 & 1.47e-03 & 9.77e-03 & 3.87e-02\\
 \hline
\end{tabular}
}
\end{center}
\end{table}

\begin{table}
\caption{Average running time of density estimation from 30 replications. Sets of 1~000 and 100~000 observations are considered. Evaluation is conducted for the entire collection of sample points in each case. Lowest average computation time for each scenario is highlighted in bold.\label{tb:runtime-n-smp-BWBS}}
\begin{center}
\scalebox{.7}{
\begin{tabular}{ll|cccccccccccc}
&&\multicolumn{8}{c}{Density}\\
\cline{3-10}
&& (a) & (b) & (c) & (d) & (e) & (f) & (g) & (h)\\ 
\hline
Exact Gauss & n = 1e+03 & 1.76e-01 & 1.79e-01 & 1.72e-01 & 1.86e-01 & 1.7e-01 & 1.83e-01 & 3.21e-01 & 2.55e-01\\
Trunc. Taylor Gauss & n = 1e+03 & 3.57e-03 & 2.9e-03 & 3.77e-03 & 3.57e-03 & 3.1e-03 & 3.63e-03 & 3.73e-03 & 3.77e-03\\
 & n = 1e+05 & 3.5e-01 & 3.2e-01 & 3.57e-01 & 3.45e-01 & 3.57e-01 & 3.53e-01 & 3.47e-01 & 3.6e-01\\
Exact $K_1$ & n = 1e+03 & {\bf 7e-04} & {\bf 9e-04} & {\bf 8.33e-04} & {\bf 7e-04} & {\bf 9e-04} & {\bf 9e-04} & {\bf 9e-04} & {\bf 7.67e-04}\\
 & n = 1e+05 & {\bf 5.85e-02} & {\bf 5.92e-02} & {\bf 5.91e-02} & {\bf 5.82e-02} & {\bf 6.1e-02} & {\bf 5.92e-02} & {\bf 5.84e-02} & {\bf 6.2e-02}\\
Exact $K_4$ & n = 1e+03 & 2.27e-03 & 2.37e-03 & 2.37e-03 & 2.27e-03 & 2.3e-03 & 2.5e-03 & 2.47e-03 & 2.33e-03\\
 & n = 1e+05 & 2.06e-01 & 2.11e-01 & 2.08e-01 & 2.04e-01 & 2.16e-01 & 2.11e-01 & 2.05e-01 & 2.16e-01\\
 \hline
\end{tabular}
}
\end{center}
\end{table}

\subsection{Density Derivative Estimation}

In this subsection the estimation of the first derivative of a density is considered. The same collection of densities used in density estimation is considered, except that density (b) is omitted since it is not differentiable at its boundaries. Of the available implementations for the methods considered, only the exact estimation and the truncated Taylor expansion for the Gaussian kernel were available. Only estimation at the sample points was considered, since all available methods are capable of this task. The average integrated squared error accuracy is reported in Table~\ref{tb:MISE-n-smp-BWBS_deriv1}. Once again the kernel $K_1$ shows the lowest error most often, however in this case only when the densities have very sharp features. The performance of the lower efficiency kernel is slightly worse on densities (a) and (d), for which the heuristic used for bandwidth selection is closer to optimal based on the AMISE objective (in the case of density (a) it is exactly optimal). In these cases the error of kernel $K_7$ is lowest.

\begin{table}
\caption{Average integrated squared error of first derivative estimates from 30 replications.. Sets of 1~000 and 100~000 observations are considered. Evaluation is conducted for the entire collection of sample points in each case. Lowest average for each scenario is highlighted in bold.\label{tb:MISE-n-smp-BWBS_deriv1}}
\begin{center}
\scalebox{.7}{
\begin{tabular}{ll|cccccccccccc}
&&\multicolumn{7}{c}{Density}\\
\cline{3-9}
&& (a) & (c) & (d) & (e) & (f) & (g) & (h)\\ 
\hline
Exact Gauss & n = 1e+03 & 4.83e-03 & 1.37e+01 & {\bf 2.39e-02} & 1.46e+01 & 8.3e+00 & 6.65e+00 & 9.91e+00\\
Trunc. Taylor Gauss & n = 1e+03 & 4.83e-03 & 1.37e+01 & 2.39e-02 & 1.46e+01 & 8.3e+00 & 6.65e+00 & 9.91e+00\\
 & n = 1e+05 & 3.71e-04 & 8.45e+00 & 2.82e-03 & 1.15e+01 & 8.06e+00 & 4.99e+00 & 8.72e+00\\
Exact $K_1$ & n = 1e+03 & 6.31e-03 & {\bf 1.24e+01} & 2.39e-02 & {\bf 1.39e+01} & {\bf 8.3e+00} & {\bf 6.21e+00} & {\bf 9.66e+00}\\
 & n = 1e+05 & 5.12e-04 & {\bf 7.35e+00} & 3.42e-03 & {\bf 1.04e+01} & {\bf 8.05e+00} & {\bf 4.2e+00} & {\bf 8.21e+00}\\
Exact $K_4$ & n = 1e+03 & 4.86e-03 & 1.42e+01 & 2.39e-02 & 1.49e+01 & 8.3e+00 & 6.95e+00 & 1e+01\\
 & n = 1e+05 & 3.75e-04 & 8.69e+00 & 2.83e-03 & 1.19e+01 & 8.07e+00 & 5.12e+00 & 8.94e+00\\
Exact $K_7$ & n = 1e+03 & {\bf 4.68e-03} & 1.49e+01 & 2.52e-02 & 1.53e+01 & 8.3e+00 & 7.12e+00 & 1.01e+01\\
 & n = 1e+05 & {\bf 3.64e-04} & 9.55e+00 & {\bf 2.82e-03} & 1.26e+01 & 8.07e+00 & 5.71e+00 & 9.19e+00\\
 \hline
\end{tabular}
}
\end{center}
\end{table}

The relative computational efficiency of the proposed approach is even more apparent in the task of density derivative estimation. Table~\ref{tb:runtime-n-smp-BWBS_deriv1} reports the average running times on all densities considered. Here it can be seen that the evaluation of the pointwise derivative at the sample points when using kernel $K_1$ is an order of magnitude faster than when using the truncated Taylor expansion. Evaluation with the kernel $K_4$ is roughly three times faster than the approximate method with the current implementations, and the running time with kernel $K_7$ is similar to the approximate approach.

\begin{table}
\caption{Average running time of estimation of first derivative from 30 replications. Sets of 1~000 and 100~000 observations are considered. Evaluation is conducted for the entire collection of sample points in each case. Lowest average computation time for each scenario is highlighted in bold.\label{tb:runtime-n-smp-BWBS_deriv1}}
\begin{center}
\scalebox{.7}{
\begin{tabular}{ll|cccccccccccc}
&&\multicolumn{7}{c}{Density}\\
\cline{3-9}
&& (a) & (c) & (d) & (e) & (f) & (g) & (h)\\ 
\hline
Exact Gauss & n = 1e+03 & 1.99e-01 & 1.6e-01 & 1.61e-01 & 1.58e-01 & 1.58e-01 & 1.59e-01 & 1.56e-01\\
Trunc. Taylor Gauss & n = 1e+03 & 5.87e-03 & 6e-03 & 5e-03 & 4.37e-03 & 5.13e-03 & 5.87e-03 & 5.63e-03\\
 & n = 1e+05 & 5.74e-01 & 5.85e-01 & 5.84e-01 & 5.64e-01 & 5.81e-01 & 5.87e-01 & 5.79e-01\\
Exact $K_1$ & n = 1e+03 & {\bf 9.67e-04} & {\bf 8.67e-04} & {\bf 9.33e-04} & {\bf 9.33e-04} & {\bf 1e-03} & {\bf 9.33e-04} & {\bf 1e-03}\\
 & n = 1e+05 & {\bf 5.82e-02} & {\bf 5.94e-02} & {\bf 5.97e-02} & {\bf 5.95e-02} & {\bf 5.87e-02} & {\bf 5.92e-02} & {\bf 5.92e-02}\\
Exact $K_4$ & n = 1e+03 & 2.1e-03 & 2.13e-03 & 2.03e-03 & 2.03e-03 & 2e-03 & 2.03e-03 & 2e-03\\
 & n = 1e+05 & 1.8e-01 & 1.83e-01 & 1.83e-01 & 1.83e-01 & 1.82e-01 & 1.83e-01 & 1.81e-01\\
Exact $K_7$ & n = 1e+03 & 3.67e-03 & 3.47e-03 & 3.67e-03 & 3.67e-03 & 3.47e-03 & 3.7e-03 & 3.5e-03\\
 & n = 1e+05 & 3.26e-01 & 3.32e-01 & 3.34e-01 & 3.33e-01 & 3.31e-01 & 3.36e-01 & 3.3e-01\\
 \hline
\end{tabular}
}
\end{center}
\end{table}

\section{Discussion and A Brief Comment on Kernel Choice}\label{sec:conclusions}

In this work a rich class of kernels was introduced whose members allow for extremely efficient and exact evaluation of kernel density and density derivative estimates.
 A much smaller sub-class was investigated more deeply. Kernels in this sub-class were selected for their simplicity of expression and the fact that they admit a large number of derivatives relative to this simplicity.
Thorough experimentation with kernels from this sub-class was conducted showing extremely promising performance in terms of accuracy and empirical running time. 

It is important to note that the efficiency of a kernel for a given task relates to the AMISE error which it induces, but under the assumption that the corresponding optimal bandwidth parameter is also selected. The popular heuristic for bandwidth selection which was used herein tends to over-estimate the AMISE optimal value when the underlying density has sharp features and high curvature. With this heuristic there is strong evidence that kernel $K_1$ represents an excellent choice for its fast computation and its accurate density estimation. On the other hand, if a more sophisticated method is employed to select a bandwidth parameter closer to the AMISE optimal, then $K_4$ is recommended for its very similar error to the popular Gaussian kernel and its comparatively fast computation.

An interesting direction for future research will be in the design of kernels in the broader class introduced herein which have simple expressions (in the sense that the polynomial component has a low degree), and which have high relative efficiency for estimation of a specific derivative of the density which is of relevance for a given task.

\bibliographystyle{Chicago}


\end{document}